# Design of LOG-MAP / MAX-LOG-MAP Decoder

**Assist.Prof. Mihai TIMIS, PhD Candidate, Dipl.Eng.**
"Gh.Asachi" Technical University of Iasi,
Automatic Control and Computer Science&Engineering Faculty
Computer Science and Engineering department
mtimis@cs.tuiasi.ro

Abstract: The process of turbo-code decoding starts with the formation of *a posteriori probabilities* (APPs) for each data bit, which is followed by choosing the data-bit value that corresponds to the *maximum a posteriori* (MAP) probability for that data bit. Upon reception of a corrupted code-bit sequence, the process of decision making with APPs allows the MAP algorithm to determine the most likely information bit to have been transmitted at each bit time.
Keywords: Turbo codes, Max-Log-Map decoder, BER, Eb/N0, convolutional codes

## 1. Introduction

This paper presents the design of the Log-Map / Max-Log-Map decoding algorithm. In figure 1 is shown the decoding algorithms available: Viterbi, SOVA, MAP, Max-Log-Map, Log-Map, improved SOVA.
    Figure 2 presents the entire Turbo Code system composed by: source encoder, channel encoder, digital modulator, transmitter, coding channel (AWGN), receiver, digital demodulator, channel decoder, source decoder, output information.

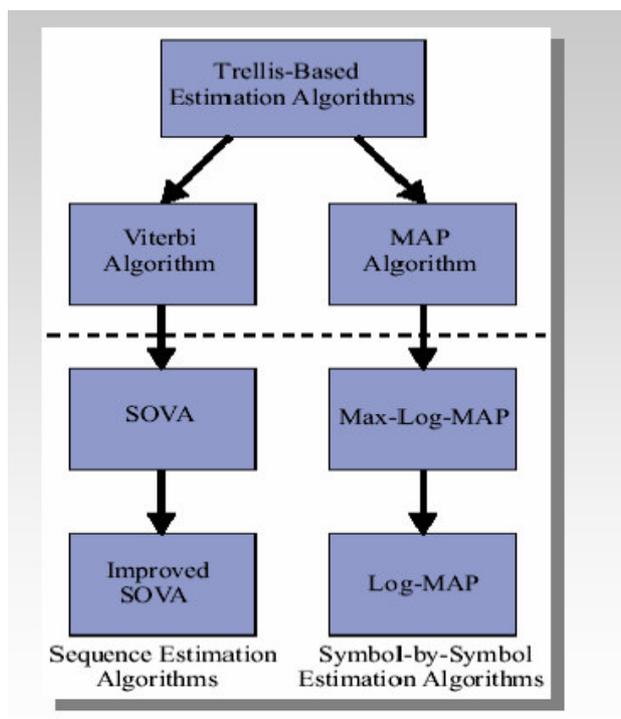

Figure 1

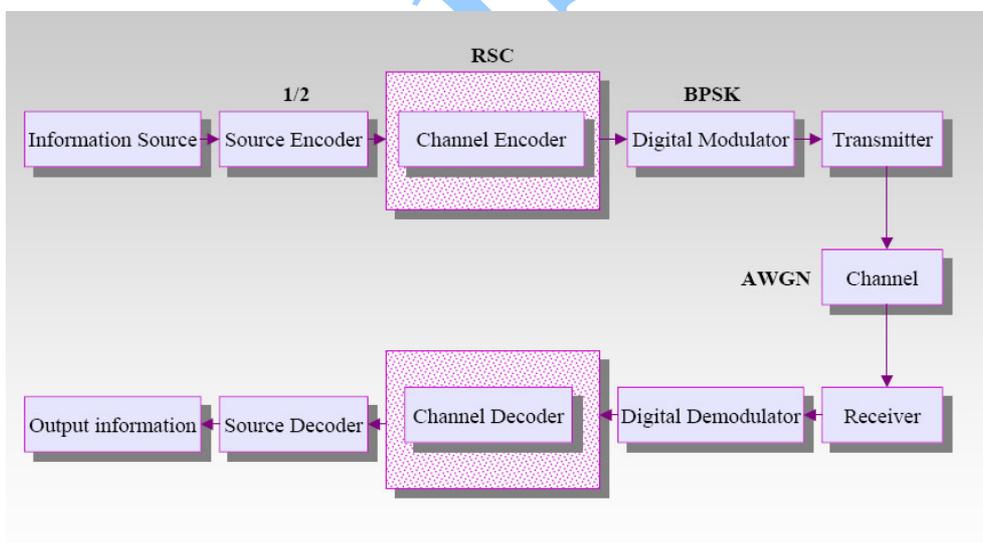

Figure 2

## 2. Principles of Iterative (Turbo) Decoding

Using Alpha, Beta and Gamma, Log-Likelihood Ratio (LLR) is computed which provides soft decision, the Soft Output makes it possible to decide if each received Bit of information is zero or one, equation (1).

$$\ln \Lambda_{t+1} = \max_{(m,m'), X=1} [\ln \gamma_{t+1}(m',m) + \ln \beta_{t+1}(m) + \ln \alpha_t(m',m)]$$
$$- \max_{(m,m'), X=-1} [\ln \gamma_{t+1}(m',m) + \ln \beta_{t+1}(m) + \ln \alpha_t(m',m)]$$
(1)

I will show an example of the performance with turbo codes. The code (1,(1+D4)/(1+D+D2+D3+D4) for both encoders but the information sequence is only transmitted from the first one. This means that the over-all rate is 1/3. The block length is 10384 bits and we use a pseudo-random interleaver. After each frame the encoders are forced to the zero state. The corresponding termination tail - 4 information bits and 4 parity bits for each encoder, a total of 16 bits - is appended to the transmitted frame and used in the decoder. In principle the termination reduces the rate, but for large frames this has no practical influence. In this case the rate is reduced from 0.3333 to 0.3332. The performance curves for Bit Error Rate (BER) and Frame Error Rate (FER) due to the sub-optimal decoding the performance curves consist of two parts. For low signal-to-noise ratios the main problem is lack of convergence in the iterated decoding process, resulting in frames with a large number of errors. In this region we are far from optimal decoding. This means that we may benefit from more iterations. As we see from the figure there is a considerable gain by going from 8 to 18 iterations, and with more iterations the performance might be even better.

In a typical communications receiver, a demodulator is often designed to produce soft decisions, which are then transferred to a decoder. The error-performance improvement of systems utilizing such soft decisions compared to hard decisions are typically approximated as 2 dB in AWGN. Such a decoder could be called a *soft input/hard output* decoder, because the final decoding process out of the decoder must terminate in bits (hard decisions). With turbo codes, where two or more component codes are used, and decoding involves feeding outputs from one decoder to the inputs of other decoders in an iterative fashion, a hard-output decoder would not be suitable. That is because hard decisions into a decoder degrades system performance (compared to soft decisions). Hence, what is needed for the decoding of turbo codes is a *soft input/soft output* decoder. For the first decoding iteration of such a soft input/soft output decoder, illustrated in Figure 3, we generally assume the binary data to be equally likely, yielding an initial a priori LLR value of $L(d) = 0$. The channel LLR value, 6 *Fundamentals of Turbo Codes Lc(x)*, is measured by forming the logarithm of the ratio of the values of $\ell 1$ and $\ell 2$ for a particular observation of *x*. The output $L(\hat{d})$ of the decoder in Figure 3 is made up of the LLR from the detector, $L'(\hat{d})$, and the extrinsic LLR output, $Le(\hat{d})$, representing knowledge gleaned from the decoding process. As illustrated in Figure 3, for iterative decoding, the extrinsic likelihood is fed back to the decoder input, to serve as a refinement of the a priori probability of the data for the next iteration.

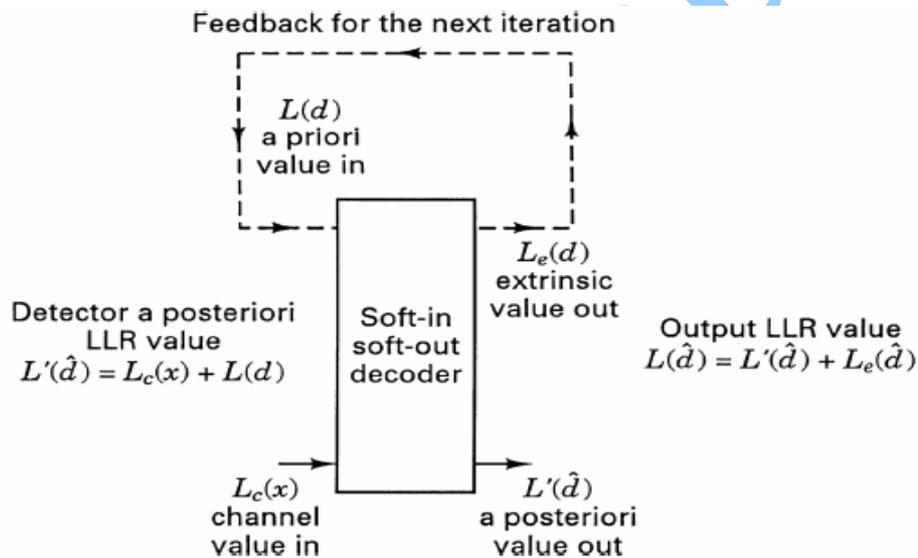

Figure 3

With the Log-Map algorithm, the Jacobi logarithm is computed exactly using the next equation which is the maximum of the function's two arguments plus a nonlinear correction function that is only a function of the absolute difference between the two arguments. The correction function $fc(|y\ 2\ x|)$ can be implemented using the log and exp functions in C (or the equivalent in other languages) or by using a large lookup table. The Log-Map algorithm is the most complex of the four algorithms when implemented in software, but as will be shown later, generally offers the best bit error rate (BER) performance. The correction function used by the Log-MAP algorithm is equation (2), along with the correction functions used by the Constant-Log-Map and Linear-Log-Map algorithms.

$$\begin{aligned} \max{}^*(x, y) &= \ln(e^x + e^y) \\ &= \max(x, y) + \ln(1 + e^{-|y-x|}) \\ &= \max(x, y) + f_c(|y - x|) \end{aligned} \qquad (2)$$

With the Max-Log-MAP algorithm, the Jacobi logarithm is loosely approximated using

$$\max{}^*(x, y) \approx \max(x, y) \qquad (3)$$

the correction function in equation (2) is not used at all. The Max-Log-Map algorithm is the least complex of the four algorithms (it has twice the complexity of the Viterbi algorithm for each half-iteration) but offers the worst BER performance. The Max-Log-Map algorithm has the additional benefit of being tolerant of imperfect noise variance estimates when operating on an AWGN channel.

## 3. Experimental data

For simulation of the Max-Log-Map algorithm (rate 1/3, frame size 483bits,1e05 informations bits) was used the Mathlab software with the Max-Log-Map toolbox. The simulation results are show in the figure 4.

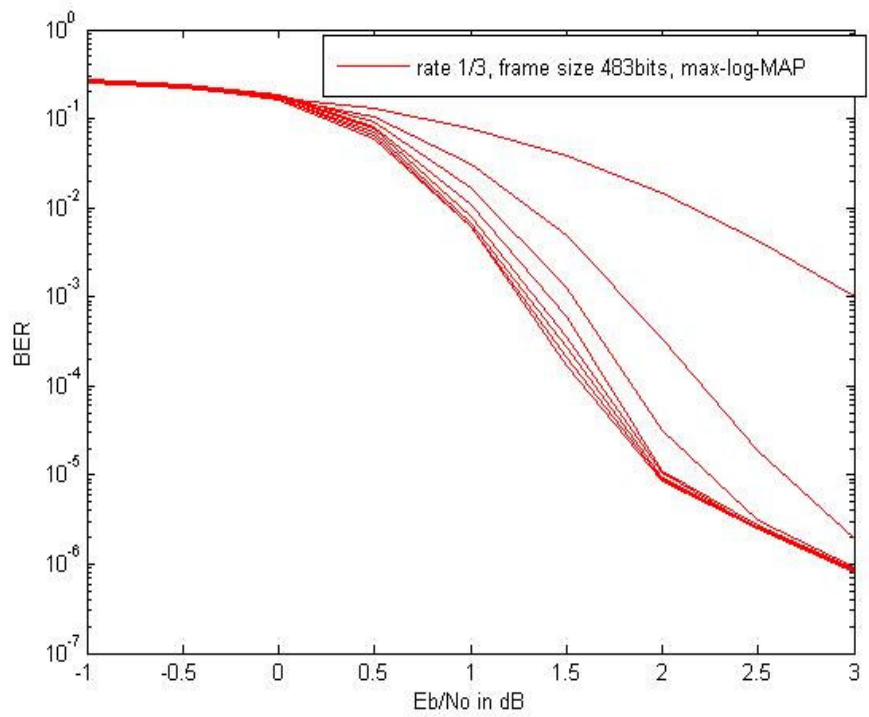

Figure 4

For simulation of the Log-Map algorithm (rate 1/3, frame size 483 bits,1e05 informations bits) was used the Mathlab software with the Log-Map toolbox. The simulation results are show in the figure 5.

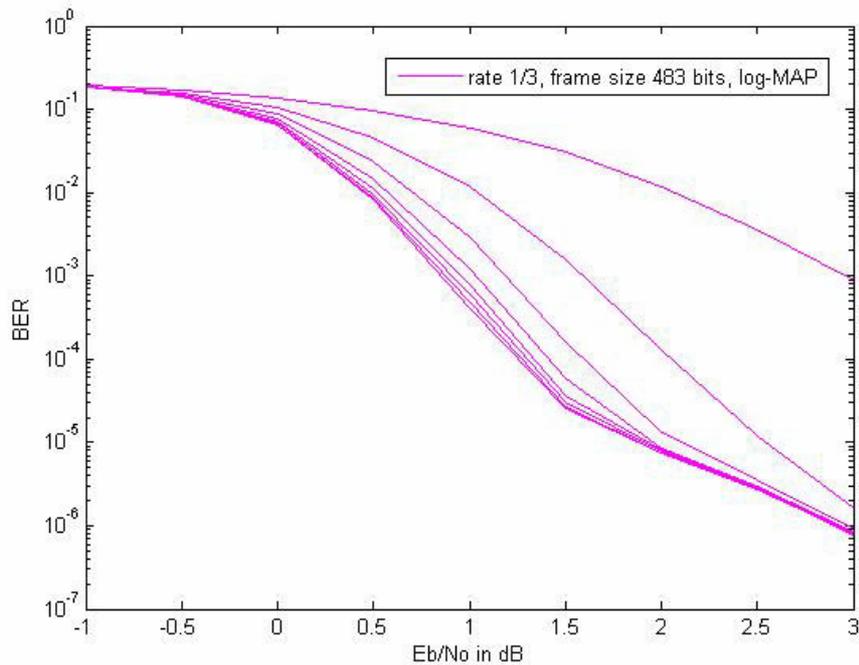

Figure 5

**References**

1. Fundamentals of turbo codes. Bernard Sklar.
2. Maximum a posteriori decoding of turbo codes. Bernard Sklar.
3. Implementation and performance of an improved Turbo Decoder on a configurable computing machine. W.Bruce Puckett. Virginia Polytechnic Institute.
4. New design of a MAP decoder. Leila Sabeti. University of Windsor, 2004.
5. A turbo tutorial. Jakob Dahl Andersen, Technical University of Denmark, 2003.
6. The UMTS Turbo Code and an efficient decoder implementation suitable for a software defined radios, M.C. Valenti, J. Sun, International Journal of Wireless Information Network, 2002.